\documentclass[aps,prd,reprint,groupedaddress,showpacs,eqsecnum,floatfix]{revtex4-1}

\usepackage{amsmath}
\usepackage{amssymb}
\usepackage{graphicx}

\begin{document}

\title{First order QED effects on a spatially flat FLRW spacetime with a Milne-type scale factor}

\author{Ion I. Cot\u aescu, Diana Popescu\\
{\small West University of Timi\c soara, V. P\^ arvan Ave. 4, RO-300223, Timi\c soara, Romania}}
%
%\date{Received: date / Revised version: date}

\begin{abstract}
The quantum electrodynamics (QED) on  a spatially flat $(1+3)$-dimensional  Friedmann-Lema\^ itre-Robertson-Walker (FLRW) spacetime with a Milne-type scale factor is outlined focusing on the amplitudes of the allowed effects in the first order of perturbations.  The definition of the transition rates is reconsidered obtaining an appropriate angular behavior of the probability of the pair creation from a photon.

\end{abstract}

\maketitle
%\tableofcontents
%\newpage
\section{Introduction}

In general relativity, the standard quantum field theory (QFT) based on perturbations is inchoate since one payed more attention to alternative non-perturbative  methods \cite{P1,P2,S1,P3,c2,c3,c4,c5} hoping to avoid some difficulties of the QFT arising even in the case of the simplest manifolds of interest in the actual cosmology, namely the spatially flat  FLRW manifolds. These manifolds, largely used in the $\Lambda CDM$ models, are symmetric under translations and, consequently, there are quantum modes expressed in terms of plane waves  with similar properties as in special relativity. For this reason these manifolds are useful for studying the behavior of the quantum matter in the presence of classical gravity turning back to the perturbation methods  of the quantum field theory where significant results were obtained by many authors \cite{Lot1,Lot2,Lot3,R1,R2,R3,AuSp1,AuSp2}.

Inspired by these results we constructed the QED in Coulomb gauge on the de Sitter expanding universe \cite{CQED}, analyzing the processes in the first order of perturbations that are allowed on this manifold since the energy and momentum cannot be conserved simultaneously \cite{Lot1,Lot2,Lot3,CQED}. Recently we completed this approach with the integral representation of the fermion propagators we need for calculating Feynman diagrams in any order of perturbations \cite{CIR}. Thus we have an example of a complete QED on a curved background but which remains singular so far.

Looking for another example  of manifold where the QED  could be constructed without huge difficulties we observed that there exists an expanding spacetime where the free field equations can be analytically  solved \cite{CIR1}. This is the $(1+3)$-dimensional spatially flat FRLW manifold whose expansion is given by a Milne-type scale factor, proportional with the proper (or cosmic) time, $t$. This spacetime is interesting and somewhat special since this is produced by gravitational sources behaving as $\frac{1}{t^2}$ in contrast with the genuine Milne's universe which is flat, having no gravitational sources \cite{BD}. 

However, the major advantage of this geometry is of hosting a friendly QED which can be constructed without major difficulties in a similar manner as the de Sitter one but giving different or even complementary results.  For example, the solutions of the free Dirac equation are similar to those on the de Sitter spacetime but with the mass and momentum changing their roles between themselves.  Thus our Milne-type spatially flat FLRW spacetime is interesting and helpful in investigating the technical features of the QED on curved backgrounds regardless the possible speculations concerning the physical interpretation of the singularity at $t=0$.   

Encouraged by these arguments we would like to investigate here the QED processes in the first order of perturbations deriving their amplitudes and studying the resulted probabilities.  We must specify, the transition rates in this new QED  must be re-defined as long as the Dirac $\delta$-function of the energy conservation is replaced by a different time integral. In this manner,  we obtain plausible probabilities but which lay out the well-known old problems of the perturbation theory, namely the infrared divergences and the divergences which appear on some particular directions  when we study the angular dependence. The first one cannot be solved at this level beefore considering a renormalization procedure. However, the angular divergences can be removed  extracting the physical results according to the method of the reduced  amplitudes proposed by Yennie and all   \cite{Yeni} and used successfully  in other investigations \cite{S3,CCS,CCS1}.  Thus we obtain new results that may be analyzed by using graphical methods.

We start in the second section defining the Milne-type spatially flat FRLW spacetime deriving the fundamental solutions of the free Dirac and Maxwell fields. In the next section we outline the QED on this spacetime deriving the first order amplitudes of the principal effects. In the fourth section we propose a new definition of the transition rate that can be applied in this case when the energy is not conserved. Thus we find that the probabilities of the transitions between charged states vanish remaining only with transitions between neutral states whose properties are studied in the next section resorting to graphical methods. Finally we present our concluding remarks.

\section{Free fields on an expandind spacetime}

Let us start defining the spacetime $M$  as the $(1+3)$-dimensional spatially flat FLRW manifold having the scale factor $a(t)=\omega t$ where $t\in (0,\infty)$ is the proper (or cosmic) time of the usual FLRW chart whose coordinates  $x^{\mu}$ (labeled  by the natural indices $\mu,\nu,...=0,1,2,3 $) are $x^0=t$ and the Cartesian space coordinates, $x^i$  ($i,j,k...=1,2,3$),  for which we may use the vector notation $\vec{x}=(x^1,x^2,x^3)$. This chart, denoted by $\{t,\vec{x}\}$, is related to the  conformal  flat one, $\{t_c,\vec{x}\}$, where we have the same space coordinates but the conformal time $t_c\in (-\infty,\infty)$ defined as
\begin{equation}\label{tt}
t_c=\int \frac{dt}{a(t)}=\frac{1}{\omega} \ln(\omega t) ~\to~  a(t_c)=e^{\omega t_c}\,.
\end{equation}
The corresponding line elements read   
\begin{eqnarray}
ds^2=g_{\mu\nu}(x)dx^{\mu}dx^{\nu}&=&dt^2-(\omega t)^2 d{\vec x}\cdot d{\vec x}\nonumber\\
&=&e^{2\omega t_c}(dt_c^2-d{\vec x}\cdot d{\vec x})\,.
\end{eqnarray}
Note that the constant $\omega$, introduced from dimensional considerations, is an useful free parameter which in the case of the genuine Milne's universe must be fixed to $\omega=1$ \cite{BD}. 

The spacetime $M$ is produced by isotropic gravitational sources, i. e. the density $\rho$ and pressure $p$, evolving in time as
\begin{equation}
\rho=\frac{3}{8\pi G}\frac{1}{t^2}\,, \quad p=-\frac{1}{8\pi G}\frac{1}{t^2}\,,
\end{equation}
and vanishing for $t\to\infty$. These sourced govern the expansion of $M$ that can be better observed in the chart $\{t, \vec{\hat x}\}$, of 'physical' space coordinates $\hat x^i=\omega t x^i$, where the line element 
 \begin{equation}
 ds^2=\left(1-\frac{1}{t^2}\vec{\hat x}\cdot \vec{\hat x}\right)dt^2 + 2 \vec{\hat x}\cdot d\vec{\hat x}\,\frac{dt}{t}-d\vec{\hat x}\cdot d\vec{\hat x}\,,
 \end{equation}
lays out an expanding horizon at $|\vec{\hat x}|=t$ and tends to the Minkowski spacetime when $t\to \infty$ and the gravitational sources vanish.

In what follows we consider only the FLRW and conformal charts where we introduce the local orthogonal non-holonomic frames defined by  the vector fields $e_{\hat\alpha}=e_{\hat\alpha}^{\mu}\partial_{\mu}$,  labeled by the local indices, $\hat\mu,\hat\nu,...=0,1,2,3$, and the associated coframes  given by the 1-forms $\omega^{\hat\alpha}=\hat e_{\mu}^{\hat\alpha}dx^{\mu}$.  In a given tetrad gauge, the  metric tensor is expressed as $g_{\mu\nu}=\eta_{\hat\alpha\hat\beta}\hat e^{\hat\alpha}_{\mu}\hat e^{\hat\beta}_{\nu}$ where $\eta={\rm diag}(1,-1,-1,-1)$ is the Minkowski metric. 
Here we are interested to chose the tertrad  gauge able to preserve the symmetry of $M$ as a global one. Bearing in mind that $M$ has the isometry group  $E(3)=T(3)\circledS SO(3)$ (of space translations and rotations) we understand that this can be done only by using  Cartesian space coordinates and  the diagonal tetrad gauge,   
\begin{eqnarray}
&e_0=\partial_t=e^{-\omega t_c}\,\partial_{t_c}\,,\quad & \omega^0=dt=e^{\omega t_c}dt_c\,, \\
&~~~e_i=\frac{1}{\omega t}\,\partial_i=e^{-\omega t_c}\,\partial_i\,, \quad & \omega^i=\omega t dx^i=  e^{\omega t_c}dx^i\,.
\end{eqnarray}
we need for writing the Dirac equation.  

For calculating Feynman diagrams we need to know the quantum modes of the free fields on $M$. Let us start with  the massive Dirac field $\psi$ of mass $m$ which  satisfy the field equation $(E_D-m)\psi =0$ where 
\begin{equation}\label{ED}
E_D=i\gamma^0\partial_{t}+i\frac{1}{\omega t}\gamma^i\partial_i
+\frac{3i}{2}\frac{1}{t}\gamma^{0}-m\,.
\end{equation}
The term of this operator depending on the Hubble function $\frac{\dot{a}}{a}=\frac{1}{t}$ can be removed at any time by substituting $\psi \to (\omega t)^{-\frac{3}{2}}\psi$. 

The fundamental solutions of the Dirac equation can be derived in the standard representation of the Dirac matrices  \cite{CIR1} or in the chiral representation (with diagonal $\gamma^0$) where we have to look for plane wave solutions of the form
\begin{eqnarray}
U_{\vec p,\sigma}(t,\vec{x})&=&[2\pi a(t)]^{-\frac{3}{2}}{e^{i\vec{p}\cdot\vec{x}}}{\cal U}_p(t) u_{\sigma}\label{U}\\
V_{\vec p,\sigma}(t,\vec{x})&=&[2\pi a(t)]^{-\frac{3}{2}}{e^{-i\vec{p}\cdot\vec{x}}}{\cal V}_p(t) v_{\sigma}
\label{V}
\end{eqnarray}
depending on  the diagonal matrix-functions
\begin{eqnarray}
{\cal U}_p(t)&=&{\rm diag}\left( u_p^+(t), u_p^-(t)\right)\,,\\
{\cal V}_p(t)&=&{\rm diag}\left(v_p^+(t), v_p^-(t)\right)\,,
\end{eqnarray}
whose matrix elements are functions only on $t$ and $p=|\vec{p}|$, determining the time modulation of the fundamental spinors. The spin part is encapsulated in  the spinors of the momentum-helicity basis that in the chiral representation of the Dirac matrices  read \cite{TH}
\begin{equation}\label{Rfspin}
u_{\sigma}=\frac{1}{\sqrt{2}}\left(
\begin{array}{l}
\xi_{\sigma}(\vec{p})\\
\xi_{\sigma}(\vec{p})
\end{array}\right) \quad
v_{\sigma}=\frac{c}{\sqrt{2}}\left(
\begin{array}{c}
-\eta_{\sigma}(\vec{p})\\
\eta_{\sigma}(\vec{p})
\end{array}\right)
\end{equation}
where  $\xi_{\sigma}(\vec{p})$ and $\eta_{\sigma}(\vec{p})= i\sigma_2 \xi_{\sigma}^{*}$ are the Pauli spinors of the helicity basis corresponding to the helicities $\sigma=\pm \frac{1}{2}$ as given in the Appendix A.  The fundamental spinors are solutions of the free Dirac equation whether  the modulation functions $u_p^{\pm}(t)$ and $v_p^{\pm}(t)$ satisfy  the first order differential equations
\begin{eqnarray}
\left(i\partial_t\pm \frac{2\sigma p}{\omega t}\right)u_p^{\pm}(t)&=&m\,u_p^{\mp}(t)\,,\label{sy1}\\
\left(i\partial_t \mp \frac{2\sigma p}{\omega t}\right)v_p^{\pm}(t)&=&-m\,v_p^{\mp}(t)\,,\label{sy2}
\end{eqnarray}
in the chart with the proper time. The solutions of these systems must satisfy the charge-conjugation symmetry \cite{CIR1},
\begin{equation}\label{VU}
v_p^{\pm}(t)=\left[u_p^{\mp}(t)\right]^*\,,
\end{equation}
and the  normalization conditions
\begin{equation}
|u_p^+|^2+|u_p^-|^2=|v_p^+|^2+|v_p^-|^2 =1\,. \label{uuvv}
\end{equation}
that determine the definitive form of the fundamental spinors, 
\begin{eqnarray}
U_{\vec{p},\sigma}(x)&=&\sqrt{\frac{mt}{\pi}}\frac{e^{i\vec{p}\cdot\vec{x}}}{[2\pi \omega t]^{\frac{3}{2}}}\left(
\begin{array}{c}
K_{\sigma-i\frac{p}{\omega}}\left(im\,t \right) \xi_{\sigma}(\vec{p})\\
K_{\sigma+i\frac{p}{\omega}}\left(im\,t \right)\xi_{\sigma}(\vec{p})
\end{array}\right)\label{U}\\
V_{\vec{p},\sigma}(x)&=&\sqrt{\frac{mt}{\pi}}\frac{e^{-i\vec{p}\cdot\vec{x}}}{[2\pi \omega t]^{\frac{3}{2}}}\left(
\begin{array}{c}
K_{\sigma-i\frac{p}{\omega}}\left(-im\,t \right) \eta_{\sigma}(\vec{p})\\
-K_{\sigma+i\frac{p}{\omega}}\left(-im\,t \right)\eta_{\sigma}(\vec{p})\\
\end{array}\right)\,,\nonumber\\ \label{V}
\end{eqnarray} 
according to the identity (\ref{H3}). Note that these spinors have a surprising structure since their modulation functions 
\begin{equation}\label{cucu}
K_{\sigma\mp i\frac{p}{\omega}}\left(im\,t \right) =K_{\frac{1}{2}\mp i\frac{2\sigma p}{\omega}}\left(im\,t \right)\,,
\end{equation}
are somewhat complementary to those derived on the de Sitter spacetime \cite{CdS}, depending on momentum through the index instead of argument. 

The fundamental spinors (\ref{U}) and (\ref{V}) form the momentum-helicity basis in which the general solutions of the Dirac equation can be expanded as
\begin{eqnarray}
&&\psi(t,\vec{x}\,) = 
\psi^{(+)}(t,\vec{x}\,)+\psi^{(-)}(t,\vec{x}\,)\nonumber\\
&& =  \int d^{3}p
\sum_{\sigma}[U_{\vec{p},\sigma}(x){\frak a}(\vec{p},\sigma)
+V_{\vec{p},\sigma}(x){\frak b}^{\dagger}(\vec{p},\sigma)]\,.\label{p3}
\end{eqnarray}
After quantization,  the particle $({\frak a},{\frak a}^{\dagger})$ and antiparticle (${\frak b},{\frak b}^{\dagger})$
operators  satisfy the canonical anti-commutation relations \cite{CIR1}, 
\begin{eqnarray}
\{{\frak a}(\vec{p},\sigma),{\frak a}^{\dagger}(\vec{p}\,\,^{\prime},\sigma^{\prime})\}&=&
\{{\frak b}(\vec{p},\sigma),{\frak b}^{\dagger}(\vec{p}\,\,^{\prime},\sigma^{\prime})\}\nonumber\\
&=&\delta_{\sigma\sigma^{\prime}}
\delta^{3}(\vec{p}-\vec{p}\,^{\prime})\,.
\end{eqnarray}
Then $\psi$ becomes a quantum free field that can be used in perturbation for calculating physical effects.  

The free Maxwell field  $A_{\mu}$ can be written easily in the conformal chart taking over the well-known results in Minkowski spacetime since the free Maxwell equations are conformally invariant. The problem is the electromagnetic gauge which does not have this property such that we are forced to adopt the Coulomb gauge with $A_0(x)=0$ as in Refs. \cite{Max,CQED},   remaining with the free Maxwell equations 
\begin{equation}\label{EMax}
\frac{1}{\sqrt{g(x)}}\,(\partial_{t_c}^2-\Delta)A_i(x)=0\,,
\end{equation}
which can be solved in momentum-helicity basis where we obtain the expansion  
\begin{equation}\label{Max}
{A_i}(x)=\int d^3k
\sum_{\lambda}\left[{\mu}_{\vec{k},\lambda;\,i}(x) \alpha({\vec
k},\lambda)+{\mu}_{\vec{k},\lambda;\,i}(x)^* \alpha^{\dagger}(\vec{k},\lambda)\right]\,,
\end{equation}
in terms of the modes functions,
\begin{equation}\label{fk}
{\mu}_{\vec{k},\lambda;\,i}(t_c,\vec{x}\,)=
\frac{1}{(2\pi)^{3/2}}\frac{1}{\sqrt{2k}}\,e^{-ikt_c+i{\vec
k}\cdot {\vec x}}\,{\varepsilon}_{i} (\vec{k},\lambda)\,,
\end{equation}
depending on the momentum $\vec{k}$ ($k=|\vec{k}|$) and helicity $\lambda=\pm 1$ of the polarization vectors ${\vec\varepsilon}_{\lambda}({\vec k})$ in Coulomb gauge (given in Appendix A). Hereby we obtain the mode functions in the FLRW chart 
\begin{equation}\label{fk1}
{\mu}_{\vec{k},\lambda;\,i}(t,\vec{x}\,)=
\frac{1}{(2\pi)^{3/2}}\frac{1}{\sqrt{2k}}(\omega t)^{-i\frac{k}{\omega}}\,e^{i{\vec
k}\cdot {\vec x}}\,{\varepsilon}_{i} (\vec{k},\lambda)\,,
\end{equation}
which will be used in our further calculations.

\section{First order QED amplitudes}

In this geometry we would like to study the QED in Coulomb gauge that can be constructed following step by step the method we used for building the  QED on the de Sitter spacetime \cite{CQED}. The principal pieces are the massive Dirac field $\psi$ and the electromagnetic potential $A_{\mu}$ minimally coupled to the gravity of $M$, interacting between themselves according to the QED action
\begin{equation}
{\cal S}=\int d^4 x\sqrt{g}\, \left[ {\cal L}_D(\psi)+{\cal L}_{M}({ A})+{\cal L}_{\rm int}(\psi,{ A})\right]\,,
\end{equation}
given by the Lagrangians of the Dirac (D) and Maxwell (M) free fields which have the standard form as in Ref. \cite{CQED}, while the interacting part, 
\begin{equation}\label{Lint}
{\cal L}_{\rm int}(\psi,{A})=-e_0
\bar{\psi}(x)\gamma^{\hat\mu}e^{\nu}_{\hat\mu}(x){A}_{\nu}(x)\psi(x)\,,
\end{equation}
corresponds to the minimal electromagnetic coupling given by the electrical charge $e_0$.

The quantization of the entire theory and the perturbation procedure based on the reduction formalism can be done just as in the de Sitter case   \cite{CQED} exploiting usual $in-out$ initial/final conditions in the conformal chart where $t_c\in (-\infty,\infty)$.  Finally, we obtain a perturbation procedure that allows us to calculate the transition amplitudes between two free states, $\alpha \to \beta$, that can be rewritten in the FLRW chart as
\begin{equation}
\langle out, \beta|in, \alpha \rangle=\langle \beta|Te^{\left(-i \int d^3x \sqrt{g}\int_{0}^{\infty}dt{\cal L}_{\rm int}\right)} |\alpha\rangle
\end{equation}
where ${\cal L}_{\rm int}$ given by Eq. (\ref{Lint})  is expressed in terms of  free fields multiplied in the chronological order  \cite{BDR}.  

Under such circumstances we can study the QED transition amplitudes since the quantum modes of the free fields on this manifold can be analytically solved. Moreover, recently we have shown that the Feynman propagators have an integral representation similar to the Fourier one used on the Minkowski spacetime  such that we can calculate Feynman diagrams of any order. However, here we restrict ourselves to the effects in the first order of the perturbation theory which are allowed because the energy is not conserved on $M$, in contrast with the Minkowski spacetime where the energy-momentum conservation forbids such effects. 

There are two types of processes involving  particles, electrons of parameters $e^-(\vec{p},\sigma)$, antiparticles,  $e^+(\vec{p}',\sigma')$ and photons $\gamma(\vec{k},\lambda)$. 
The first type is of the processes whose  $in$ and $out$ states are charged as, for example,  in the case of the photon adsorption $e^-+\gamma \to e^-$ whose amplitude reads
\begin{eqnarray}
&&A_{\sigma'}^{\sigma,\lambda}(\vec{p},\vec{k};\vec{p}')=\langle e^-(\vec{p}',\sigma') |{\bf S}_1|e^-(\vec{p},\sigma),\gamma(\vec{k},\lambda)\rangle\nonumber\\
&&=-ie_0\int{d^4x}{(\omega
t)^2}\,\overline{U}_{{\vec{p}\,}',\sigma'}(x)\, {\gamma^i}
{\mu}_{\vec{k},\lambda;\,i}(x) U_{\vec{p},\sigma}(x)\,.\label{Amp1}
\end{eqnarray}
When the photon is adsorbed by a positron we have to replace
${U}_{{\vec{p}\,}',\sigma'}\to {V}_{\vec{p},\sigma}$ and
${U}_{\vec{p},\sigma}\to {V}_{{\vec{p}\,}',\sigma'}$. Moreover, if
we replace ${\mu_i}\to \mu_i^*$ then we obtain the amplitudes of
the transitions $e^-\to e^- +\gamma$ and respectively $e^+\to e^++\gamma$ in
which a photon is emitted.

The second type of amplitudes involves only neutral $in$ and $out$ states as in the cases of the pair creation, $\gamma \to e^- + e^+$, and annihilation, $e^- + e^+ \to \gamma$, when we
find the related amplitudes
\begin{eqnarray}
&&A^{\lambda}_{\sigma,\sigma'}(\vec{k};\vec{p},\vec{p}')=\langle e^-(\vec{p},\sigma),e^+(\vec{p}',\sigma') |{\bf S}_1|\gamma(\vec{k},\lambda)
\rangle \nonumber\\
&&\hspace*{21mm}\,=-\langle \gamma(\vec{k},\lambda)|{\bf S}_1|e^- (\vec{p},\sigma),
e^+(\vec{p}',\sigma') \rangle^*\nonumber\\
&&=-ie_0\int{d^4x}{(\omega
t)^2}\,\overline{U}_{\vec{p},\sigma}(x)\, {\gamma^i}
{\mu}_{\vec{k},\lambda;\,i}(x) V_{{\vec{p}\,}',\sigma'}(x)\,.\label{Amp2}
\end{eqnarray}
If we replace ${\mu}_i\to {\mu}_i^{*}$ in Eq. (\ref{Amp2}) then we obtain the amplitudes of the creation of leptons from vacuum, $vac\to e_++e_-+\gamma$ or their annihilation to vacuum, $e_++e_-+\gamma\to vac$.

In what follows we focus on the amplitudes  (\ref{Amp1}) and (\ref{Amp2}) that can be calculated by using the previous results and taking into account that we work with the chiral representation of the Dirac matrices. Thus we obtain
\begin{eqnarray}
&&A^{\sigma,\lambda}_{\sigma'}(\vec{p},\vec{k};\vec{p}')=i \frac{e_0 m}{\pi}\frac{\omega^{-i\frac{k}{\omega}-1}}{\sqrt{2 k}\,(2\pi)^{\frac{3}{2}}}\nonumber\\
&&~~~\times \delta^3(\vec{p}+\vec{k}-\vec{p}') \,\Pi_{\sigma'}^{\sigma,\lambda}(\vec{p},\vec{k};\vec{p}')\,I^-_{\sigma',\sigma}({p'},{p},{k})\,,\label{Aa1}\\
&& A^\lambda_{\sigma,\sigma'}(\vec{k};\vec{p},\vec{p}')=i \frac{e_0 m}{\pi}\frac{\omega^{-i\frac{k}{\omega}-1}}{\sqrt{2 k}\,(2\pi)^{\frac{3}{2}}}\nonumber\\
&&~~~\times \delta^3(\vec{p}+\vec{p}'-\vec{k}) \, \Pi_{\sigma,\sigma'}^{\lambda}(\vec{k};\vec{p},\vec{p}')\,I^+_{\sigma,\sigma'}({p},{p}',{k})\,,\label{Ac1}
\end{eqnarray}
where we separate the terms depending on polarizations,
\begin{eqnarray}
 \Pi_{\sigma'}^{\sigma,\lambda}(\vec{p},\vec{k};\vec{p}')&=&\xi_{\sigma'}^+(\vec{p}')\sigma_i\varepsilon_i(\vec{k},\lambda)\xi_{\sigma}(\vec{p})\,,\\
 \Pi_{\sigma,\sigma'}^{\lambda}(\vec{k};\vec{p},\vec{p}')&=&\xi_{\sigma}^+(\vec{p})\sigma_i\varepsilon_i(\vec{k},\lambda)\eta_{\sigma'}(\vec{p}')\,,
\end{eqnarray}
from the time integrals 
\begin{equation}
I^{\pm}_{\sigma,\sigma'}({p},{p}',{k})=\int_{0}^{\infty} dt\,{\cal K}^{\pm}_{\sigma,\sigma'}({p},{p}',{k};t)\,,
\end{equation}
whose time-dependent functions
\begin{eqnarray}\label{Kf}
&&{\cal K}^{\pm}_{\sigma,\sigma'}({p},{p}',{k};t)=t^{ i\frac{k}{\omega}}\left[K_{\sigma+i\frac{p}{\omega}}(- imt)K_{\sigma'-i\frac{p'}{\omega}}(\mp imt)\right. \nonumber\\
&&\left.\hspace*{16mm}\pm K_{\sigma-i\frac{p}{\omega}}(- imt)K_{\sigma'+i\frac{p'}{\omega}}(\mp imt)\right]\,,
\end{eqnarray}
result from Eqs. (\ref{U}) and (\ref{V}).

These integrals have remarkable properties,
\begin{eqnarray}
&&I^{\pm}_{\sigma,\sigma'}({p},{p}',{k})=\pm I^{\pm}_{-\sigma,-\sigma'}({p},{p}',{k})=\pm I^{\pm}_{\sigma,\sigma'}({-p},{-p}',{k})\nonumber\\
&&~~~~~~~=I^{\pm}_{\sigma,-\sigma'}({p},{-p}',{k})=I^{\pm}_{-\sigma,\sigma'}({-p},{p}',{k})\,,
\end{eqnarray}
since $K_{\nu}=K_{-\nu}$, and  can be solved according to Eq. (\ref{KKKK}) obtaining, after a few manipulations, the following quantities we need for deriving the transition probabilities: 
\begin{eqnarray}
&&\left|I^+_{\pm\frac{1}{2},\pm\frac{1}{2}}(p,p',k)\right|=\Delta(p,p',k)\,e^{\frac{\pi k}{2\omega}}\,,\label{Ipp}\\
&&\left|I^+_{\mp\frac{1}{2},\pm\frac{1}{2}}(p,p',k)\right|=\Delta(p,-p',k)\,e^{\frac{\pi k}{2\omega}}\,,\label{Imp}\\
&&\left|I^-_{\pm\frac{1}{2},\pm\frac{1}{2}}(p,p',k)\right|=\Delta(p,p',k)\,e^{\frac{\pi k}{2\omega}}\nonumber\\
&&~~~~~~\times  \left|\sinh \frac{\pi p}{\omega}\pm\frac{p'-p}{k}\cosh\frac{\pi p}{\omega}\right|\,, \\
&&\left|I^-_{\mp\frac{1}{2},\pm\frac{1}{2}}(p,p',k)\right|=\Delta(p,-p',k)\,e^{\frac{\pi k}{2\omega}}\nonumber\\
&&~~~~~~\times  \left|\sinh \frac{\pi p}{\omega}\pm\frac{p+p'}{k}\cosh\frac{\pi p}{\omega}\right|\,,
\end{eqnarray}
where
\begin{eqnarray}
&&\Delta(p,p',k) =\frac{\pi^{\frac{3}{2}}\sqrt{\omega}}{2m}\,
\left[\frac{k\,{\rm sinh}\frac{k\pi}{\omega}}{k^2-(p-p')^2}\right]^{\frac{1}{2}} \nonumber\\   
&&\times\left[ {\rm sinh}\left(\frac{\pi(k-p+p')}{2\omega} \right)  {\rm sinh}\left(\frac{\pi(k+p-p')}{2\omega} \right) \right.\nonumber\\
&&\left.\times\, {\rm cosh}\left(\frac{\pi(k+p+p')}{2\omega} \right) {\rm cosh}\left(\frac{\pi(k-p-p')}{2\omega} \right)\right]^{-\frac{1}{2}}\,.\nonumber\\\label{Delta}
\end{eqnarray} 
We observe that the function $\Delta(p,p',k)$ satisfies
\begin{equation}
\Delta(p,p',k)=\Delta(-p,-p',k)=\Delta(p,p',-k)\,,
\end{equation}
being  singular for $k\pm(p-p')=0$. Note that the function  $\Delta(p,-p',k)$ is singular only for $k=(p+p')$ since $k,p,p'\in {\Bbb R}^+$. 

\section{Rates and probabilities}

The results of the previous section lead to the conclusion that the amplitudes of the transitions $\alpha\to \beta$ have the general form
\begin{equation}\label{Aab}
A_{\alpha\beta}
=\langle out\, \beta|in\, \alpha\rangle=\delta^3(\vec{p}_{\alpha}-\vec{p}_{\beta}) M_{\alpha\beta}I_{\alpha\beta}\,,
\end{equation}
laying out the Dirac $\delta$-function of the momentum conservation but without conserving the energy.  Thus the time integration gives the quantity
\begin{equation}
I_{\alpha\beta}=\int_0^{\infty}dt\, {\cal K}_{\alpha\beta}(t)\,,
\end{equation} 
 instead of the familiar $\delta(E_{\alpha}-E_{\beta})$ we meet in the flat case when the energy is conserved. This could lead to some difficulties when we calculate the transition probabilities.

We remind the reader that  in the usual QED on Minkowski spacetime the transition probabilities are derived from amplitudes satisfying the energy-momentum conservation,
\begin{equation}
\hat A_{\alpha\beta}
=\delta(E_{\alpha}-E_{\beta})\delta^3(\vec{p}_{\alpha}-\vec{p}_{\beta})\hat M_{\alpha\beta}\,,
\end{equation}
evaluating $\delta(0)\delta^3(0)\sim\frac{1}{(2\pi)^4}TV$ in terms of the total volume $V$ and interaction time $T$ such that one obtains the probability per unit of volume and unit of time as \cite{LL,BDR}
\begin{equation}
\hat{\cal P}_{\alpha\beta}=\frac{|\hat A_{\alpha\beta}|^2}{VT}=\delta(E_{\alpha}-E_{\beta})\delta^3(\vec{p}_{\alpha}-\vec{p}_{\beta})\frac{| \hat M_{\alpha\beta}|^2}{(2\pi)^4}\,.
\end{equation} 
In fact this is the transition rate per unit of volume we refer here simply as rate denoted by ${\cal R}$.

In our QED on $M$ the rates must be derived in another manner since the amplitudes have here different forms as in Eq. (\ref{Aab}). Therefore, we introduce first the time-dependent amplitudes 
\begin{eqnarray}
A_{\alpha\beta}(t)
&=&\delta^3(\vec{p}_{\alpha}-\vec{p}_{\beta}) M_{\alpha\beta}I_{\alpha\beta}(t)\nonumber\\
&=&\delta^3(\vec{p}_{\alpha}-\vec{p}_{\beta}) M_{\alpha\beta}\int_0^t dt'{\cal K}_{\alpha\beta}(t')\,,
\end{eqnarray} 
that can be rewritten in terms of the conformal time as $A_{\alpha\beta}(t_c)=A_{\alpha\beta}[t(t_c)]$.  Then we define the transition rate according to Eq. (\ref{tt}) as
\begin{equation}\label{Rat}
{\cal R}_{\alpha\beta}=\lim_{t_c\to\infty} \frac{1}{2V}\frac{d}{dt_c}\left|A_{\alpha\beta}(t_c)\right|^2=\lim_{t\to\infty} \frac{\omega t}{2V}\frac{d}{dt}\left|A_{\alpha\beta}(t)\right|^2
\end{equation}   
obtaining the final result
\begin{equation}
{\cal R}_{\alpha\beta}=\delta^3(\vec{p}_{\alpha}-\vec{p}_{\beta})\frac{|M_{\alpha\beta}|^2}{(2\pi)^3}|I_{\alpha\beta}| K_{\alpha\beta}
\end{equation}
where
\begin{equation}\label{lim}
K_{\alpha\beta}=\lim_{t\to \infty}\left|\omega t \,{\cal K}_{\alpha\beta}(t)\right|\,.
\end{equation}
Note that the basic definition (\ref{Rat}) is given in the conformal chart where the $in$ and $out$ states can be defined correctly in the domain $-\infty<t_c<\infty$, as in the flat case or in our de Sitter QED. 

Thus for calculating the transition rates of the processes under consideration here we need to calculate the limits (\ref{lim}) of the functions (\ref{Kf}). Fortunately, this can be done easily since the modified Bessel functions have a simple asymptotic behavior as in Eq. (\ref{Km0}).
Thus we obtain the dramatic result, 
\begin{eqnarray}
\lim_{t\to \infty}\omega t\left|{\cal K}^{+}_{\sigma,\sigma'}({p},{p}',{k};t) \right|&=&\frac{\pi\omega}{m}\,,\\
\lim_{t\to \infty}\omega t\left|{\cal K}^{-}_{\sigma,\sigma'}({p},{p}',{k};t) \right|&=&0\,,
\end{eqnarray}
which shows that the rates of all the processes involving charged states vanish, remaining only with the transitions between neutral states  For the transition, $\gamma(\vec{k},\lambda)\to e^-(\vec{p},\sigma)+e^+(\vec{p}',\sigma')$ we obtain  the expression
\begin{eqnarray}
&&{\cal R}^{\lambda}_{\sigma,\sigma'}(\vec{k};\vec{p},\vec{p}')=\frac{e_0^2}{(2\pi)^7}\,\frac{m\omega}{k}\,\delta^3(\vec{p}+\vec{p}'-\vec{k})\nonumber\\
&&~~~~~~\times |\Pi_{\sigma,\sigma'}^{\lambda}(\vec{k};\vec{p},\vec{p}')|^2\,|I^+_{\sigma,\sigma'}({p},{p}',{k})|\,,
\end{eqnarray}
which allows us to derive the probability per units of volume and time  integrating over $\vec{k}$. Thus we obtain  
\begin{eqnarray}
&&{\cal P}^{\lambda}_{\sigma,\sigma'}(\vec{p},\vec{p}')=\int \frac{d^3k}{(2\pi)^3}\, {\cal R}^{\lambda}_{\sigma,\sigma'}(\vec{k};\vec{p},\vec{p}')=\frac{e_0^2}{(2\pi)^{10}}\,\frac{m\omega}{k(\theta)}\nonumber\\
&&~~~~~~\times |\Pi_{\sigma,\sigma'}^{\lambda}(\vec{p}+\vec{p}';\vec{p},\vec{p}')|^2\,|I^+_{\sigma,\sigma'}({p},{p}',{k}(\theta))|\,,
\end{eqnarray}
where
\begin{equation}
k(\theta)=\left|\vec{p}+\vec{p}'\right|=\sqrt{p^2 +2p p'\cos\theta +{p'}^2}
\end{equation}
depend on the angle $\theta$ between $\vec{p}$ and $\vec{p}'$.

{ \begin{figure}
 \centering
    \includegraphics[scale=0.50]{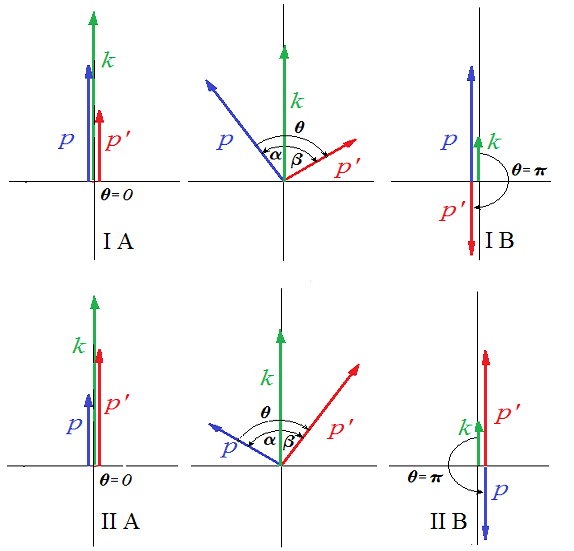}
    \caption{Pair production in the frame $\{e\}$ (I) for $p>p'$: (I A)   $\theta=0$ $\to$ $k=p+p'$,  $\sigma'=\sigma$ and $\lambda=2\sigma$, (I B) $\theta=\pi$ $\to$ $k=p'-p$,  $\sigma'=-\sigma$ and $\lambda=2\sigma$, and (II) for $p<p'$ :  (II A)  $\theta=0$ $\to$ $k=p+p'$,  $\sigma'=\sigma$ and $\lambda=2\sigma$, (II B) $\theta=\pi$ $\to$ $k=p'-p$ and $\sigma'=-\sigma$ and $\lambda=-2\sigma$.  }
  \end{figure}}

For studying these probabilities we have to calculate the polarization terms which are extremely complicated in an arbitrary geometry. For this reason it is  convenient to consider a particular frame $\{e\}=\{\vec{e}_1,\vec{e}_2,\vec{e}_3\}$ in the momentum space where  $\vec{k}=\vec{p}+\vec{p}'=k(\theta) \vec{e}_3$ and the vectors $\vec{p}$ and $\vec{p}'$ are in the plane $\{\vec{e}_1,\vec{e}_3\}$ (as in Fig. 1)  having the spherical coordinates  $\vec{p}=(p,\alpha,0)$ and ${\vec{p}\,}'=(p',\beta,\pi)$ such that  
\begin{eqnarray}
\theta&=&\alpha+\beta\,,\label{theta1}\\
p\sin\alpha&=&p'\sin\beta\,.\label{pp1}
\end{eqnarray}
In this geometry the polarization vectors take the simple form 
$\vec{\varepsilon}_{\pm 1}(\vec{k})=\frac{1}{\sqrt{2}}(\pm
\vec{e}_1-i\vec{e}_2)$ that allows us to derive the polarization matrices
\begin{eqnarray}
\hat \Pi^{\lambda=1}&=&\sqrt{2}\,\left(
\begin{array}{cc}
\cos\frac{\alpha}{2}\cos\frac{\beta}{2}&\cos\frac{\alpha}{2}\sin\frac{\beta}{2}\\
\sin\frac{\alpha}{2}\cos\frac{\beta}{2}&\sin\frac{\alpha}{2}\sin\frac{\beta}{2}
\end{array}\right)\,,\label{matix1a}\\
\hat \Pi^{\lambda=-1}&=&\sqrt{2}\,\left(
\begin{array}{cc}
\sin\frac{\alpha}{2}\sin\frac{\beta}{2}&\sin\frac{\alpha}{2}\cos\frac{\beta}{2}\\
\cos\frac{\alpha}{2}\sin\frac{\beta}{2}&\cos\frac{\alpha}{2}\cos\frac{\beta}{2}
\end{array}\right)\,,\label{matix2a}
\end{eqnarray}
whose matrix elements, $\left|{\hat \Pi}^{\lambda}_{\sigma,\sigma'}\right|$ are the absolute values of the polarization terms in the particular frame $\{e\}$. 
Now we can choose the free parameters $p,\,p'$ and $\theta$ since the angles we need for calculating the polarization matrix can be deduced as
\begin{eqnarray}
\alpha&=&{\rm arctan}\left(\frac{p'\sin \theta}{p+p'\cos\theta}\right)\,,\\
\beta&=&\theta-{\rm arctan}\left(\frac{p'\sin \theta}{p+p'\cos\theta}\right)\,,
\end{eqnarray}
when $p>p'$, as it results from Eqs. (\ref{theta1}) and (\ref{pp1}). 
For $p<p'$ we obtain similar relations changing $\alpha \leftrightarrow \beta$ and $p \leftrightarrow p'$ while for $p=p'$ we have $\alpha=\beta=\frac{\theta}{2}$ . 

{ \begin{figure}
 \centering
    \includegraphics[scale=0.40]{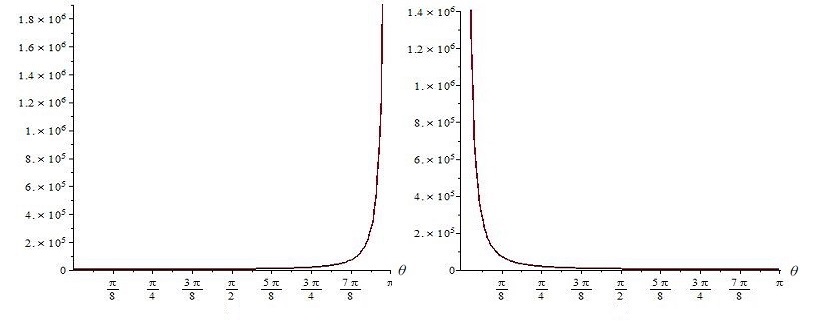}
    \caption{The singular behavior of the functions $\Delta(p,p',k(\theta))$ (left panel) and $\Delta(p,-p',k(\theta))$ (right panel) for $p=0.01\,\omega$ and $p'=0.03\,\omega$.}
  \end{figure}}
  
Finally, according to Eqs. (\ref{Ipp}) and (\ref{Imp}) we obtain the definitive result in the frame $\{e\}$ where the probability per unit of volume and unit of  time, 
\begin{eqnarray}
{\cal P}^{\lambda}_{\sigma,\sigma'}(p,p',\theta)&=&\frac{e_0^2}{(2\pi)^{10}}\,\frac{m\omega}{k(\theta)}\, e^{\frac{\pi k(\theta)}{2\omega}}\nonumber\\
&\times&\hat\Pi_{\sigma,\sigma'}^2\,\Delta(p,{\rm sign}(\sigma\sigma')p',k(\theta))\,,\label{Pfinal}
\end{eqnarray}
depends only on polarization and the free parameters $(p,p',\theta)$ through the polarization term and  the function  $\Delta(p,p',k)$ defined by Eq. (\ref{Delta}).  

\section{Graphical analysis}

The closed expression (\ref{Pfinal}) encourages us to study the dependence of this probability on the kinematic parameters and polarizations. We observe first that here we cannot speak  about the polarization conservation since we work in the momentum-helicity basis. Nevertheless, there are some particular positions in which the momenta have the same direction and, consequently, the polarizations must be conserved as spin projections on the same direction. These positions are obtained either for $\theta=0$, as in the panels I A and II A of Fig. 1, when we have 
\begin{equation}\label{Sel1}
\alpha=\beta=0 \to  p'=p+k\,, \quad \lambda=\sigma+\sigma'\,,
\end{equation} 
or for $\theta=\pi$ when we find two different cases presented in the panels II A and respectively II B.  In the first one (I B) we set $p>p'$ and consequently
\begin{equation}\label{Sel2}
\alpha=0\,,\beta=\pi \to k=p-p'\,,\quad \lambda=\sigma-\sigma'\,,
\end{equation}
while in the second one (II B) the situation is reversed such that  $p<p'$ and
\begin{equation}\label{Sel3}
 \alpha=\pi\,,\beta=0 \to k=p'-p\,, \quad \lambda=\sigma'-\sigma\,.
\end{equation}
Note that when $p=p'$ we remain only with the parallel case (I A=II A)   since the anti-parallel equal momenta lead to $k=0$ when the photon of the $in$ state disappears. 

{ \begin{figure}
 \centering
    \includegraphics[scale=0.38]{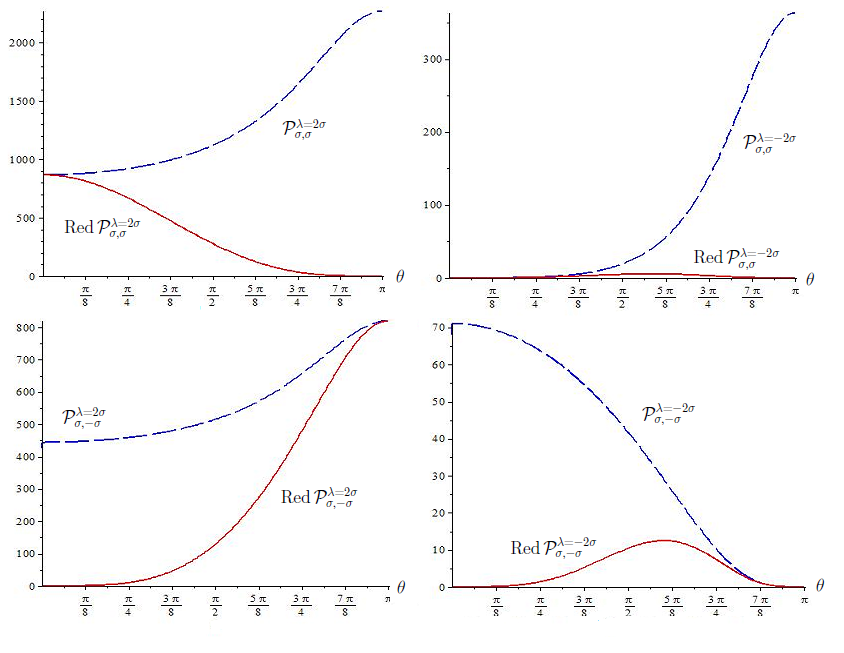}
    \caption{The effect of the reduction procedure: the original (dashed lines) and  reduced  (solid lines)  probabilities versus $\theta$ for different polarizations and $p=0.05\, \omega$ and $p'=0.02\,\omega$.}
  \end{figure}}

Now we expect to recognize the above selection rules by plotting the probabilities (\ref{Pfinal}) versus $\theta$ for fixed values of the momenta $p$ and $p'$.  The unpleasant surprise is of finding a wrong behaviors just for the angles $\theta=0$ or  $\theta=\pi$ for which the selection rules require the probabilities to vanish if the polarizations are not conserved. This is because of the function $\Delta(p,p',k(\theta))$ which becomes singular   
for $k\pm(p-p')=0$ having the profile plotted in Fig. 2.

Thus we meet again the sickness of the perturbation procedures leading to singularities or violation of the conservation rules on some particular directions. In order to extract the physical information we need to remove these effects resorting to the method of Yennie and all \cite{Yeni} of constructing the reduced  amplitudes by multiplying the calculated one by suitable trigonometric functions.  Thus, for example, the singularity at $\theta=0$ of the scattering amplitudes of various scattering processes can be removed by multiplying the amplitude with $(1-\cos\theta)^n$ where $n$ gives the reduction order.  In the case of our amplitudes the reduction of the first order, with $n=1$, is enough for eliminating the singularities in $\theta=0$ and $\theta=\pi$  if we define the reduced probabilities as
\begin{eqnarray}
{\rm Red}\, {\cal P}^{\lambda=\pm 2\sigma}_{\sigma,\sigma}(p,p',\theta)&=&{\cal P}^{\lambda=\pm 2\sigma}_{\sigma,\sigma}(p,p',\theta) \cos^4\frac{\theta}{2}\,,\\
{\rm Red}\, {\cal P}^{\lambda=\pm 2\sigma}_{\sigma,-\sigma}(p,p',\theta)&=&{\cal P}^{\lambda=\pm 2\sigma}_{\sigma,-\sigma}(p,p',\theta) \sin^4\frac{\theta}{2}\,.
\end{eqnarray}
Now we can verify that these  match perfectly with the selection rules (\ref{Sel1})-(\ref{Sel3}) by plotting them on the whole domain $\theta\in [0,\pi]$ as in Figs. 3 and 4. Moreover, we observe that the reduction procedure does not affect the physical content since for the angles $\theta=0$ and $\theta=\pi$ for which the function $\Delta$ is regular we have ${\rm Red}\, {\cal P}^{\lambda}_{\sigma,\sigma'}={\cal P}^{\lambda}_{\sigma,\sigma'}$ as we see in Fig. 3. Thus we can conclude that the reduction procedure is correct helping us to understand the physical behavior of the analyzed process.

{ \begin{figure}
 \centering
    \includegraphics[scale=0.38]{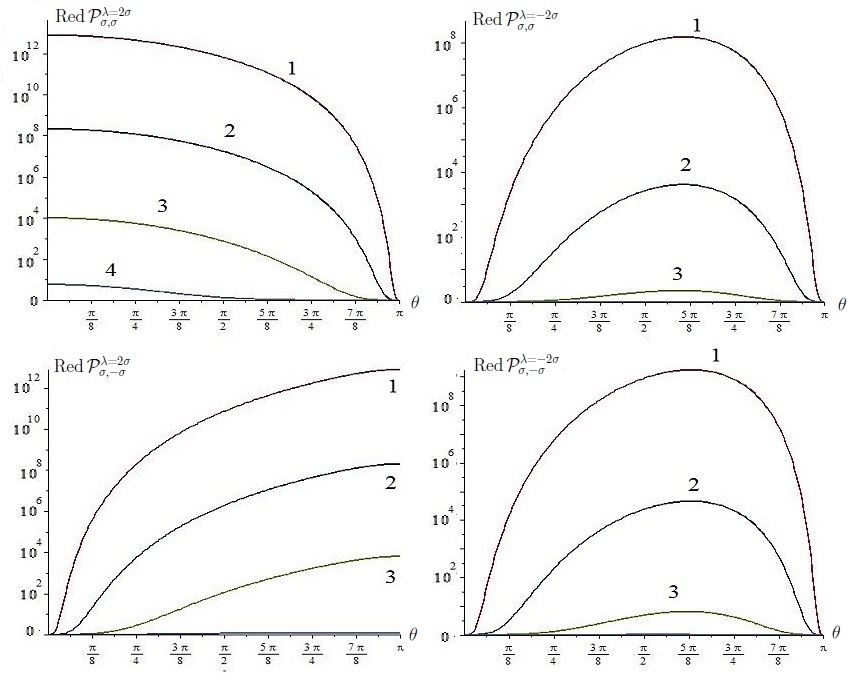}
    \caption{Reduced probabilities versus $\theta$ for different polarizations and momenta: (1) $p=0.002\, \omega$ and $p'=0.001\,\omega$ (2) $p=0.02\, \omega$ and $p'=0.01\,\omega$. (3) $p=0.2\, \omega$ and $p'=0.1\,\omega$ (4) $p=2\, \omega$ and $p'=\omega$}
  \end{figure}}  

On the other hand, we must specify that another problem is the divergence at $p\sim p'=0$. Indeed,  as we see in Fig. 4, the reduced probabilities increase when the momenta $p$ and $p'$ are decreasing such that for vanishing momenta the probabilities diverge,
\begin{equation}
\lim_{\substack{p\to 0\\p'\to 0}} {\cal P}^{\lambda}_{\sigma,\sigma'}=\lim_{\substack{p\to 0\\p'\to 0}}{\rm Red}\,{\cal P}^{\lambda}_{\sigma,\sigma'} = \infty\,.
\end{equation}
This unwanted effect is somewhat analogous to the infrared catastrophe of the usual QED and 
could be of interest in a future procedure of the vertex renormalization. 

Finally we note that the dependence on the parameter $\omega$ is almost trivial since for large values of $\omega$ the probabilities are increasing linearly with this parameter.  
  
\section{Concluding remarks}

We visited here for the first time the world of the quantum fields on the spatially flat FLRW spacetime with a Milne-type modulation factor (denoted here by $M$). The first impression was that this manifold, born from a time singularity, might produce new spectacular physical effects but, in fact, our calculations show that, at least from the point of view of the quantum theory, this spacetime behaves normally producing similar  effects as the de Sitter expanding universe \cite{CQED}.  The only notable new feature is that the first order transitions between charged states are forbidden but we cannot say if this is specific to this geometry as long as we do not have other examples.

From the technical point of view, $M$ and  the de Sitter spacetime have complementary behaviors  as we can see from the next self-explanatory table, 
\begin{center}
\begin{tabular}{lcc}
&$M$&de Sitter\\
&&\\
$t$&$0<t=\frac{1}{\omega}e^{\omega t_c}<\infty$& $-\infty<t<\infty$\\
$t_c$&$-\infty<t_c<\infty$&$-\infty<t_c=-\frac{1}{\omega}e^{-\omega t}<-\frac{1}{\omega}$ \\
$a(t)$&$\omega t$&$e^{\omega t}~~~~~$\\
$a(t_c)$&$e^{\omega t_c}$&$-\frac{1}{\omega t_c}$\\
$u^{\pm}_p$&$K_{\frac{1}{2}\mp i \,\frac{2\sigma p}{\omega}}( im t)$&$K_{\frac{1}{2}\mp i\frac{m}{\omega}}( i p t_c)$
\end{tabular}
\end{center}
where we used the identity (\ref{cucu}) denoting by $\omega$ the free parameter of $M$ and the Hubble constant of the de Sitter expanding portion \cite{CdS}. Thus we have at least two related examples that will help us to construct the perturbative QFT on curved backgrounds.

In other respects, it is worth noting that we met here the same general problems of the  scattering theory known from long time on the flat spacetime namely, the infrared catastrophe and a divergent angular behavior  on special directions as the forward and backward ones. We have seen that the angular divergences can be removed by using a reduction procedure \cite{Yeni} but we cannot comment about the infrared divergences in this stage of the theory when the processes in the second or higher orders of perturbations are not yet calculated and, consequently,  we have no idea about the renormalization procedures that could be applied in this case.   

We hope that the study of  these problems will lead to a better understanding of the perturbative QFT on the FLRW spacetimes and even to a possible integration of the perturbative and non-perturbative methods in a future general and effective new QFT.

\appendix
\section{Polarization}

The Pauli spinors of the momentum-helicity basis, $\xi_{\sigma}(\vec{p})$, of helicity $\sigma =\pm\frac{1}{2}$,   satisfy the eigenvalues problem $(\vec{p}\cdot \vec{S})\,\xi_{\sigma}(\vec{p})=\sigma\, p\, \xi_{\sigma}(\vec{p})$ where $S_i=\frac{1}{2}\sigma_i$ are the spin operators expressed in terms of Pauli matrices.  They have the form 
\begin{eqnarray}
\xi_{\frac{1}{2}}(\vec{p})&=&\sqrt{\frac{p+p^3}{2p}}\left(
\begin{array}{c}
1\\
\frac{p^1+i p^2}{p+p^3}
\end{array}\right)\,,\\ 
\xi_{-\frac{1}{2}}(\vec{p})&=&\sqrt{\frac{p+p^3}{2p}}\left(
\begin{array}{c}
\frac{-p^1+i p^2}{p+p^3}\\
1
\end{array}\right)\,.
\end{eqnarray}
The antiparticle spinors are defined usually as $\eta_{\sigma}(\vec{p})=i\sigma_2 \xi_{\sigma}(\vec{p})^*$  \cite{BDR,TH} in order to satisfy $(\vec{p}\cdot \vec{S})\,\eta_{\sigma}(\vec{p})=-\sigma\, p\, \eta_{\sigma}(\vec{p})$. 

The polarization of the free Maxwell field is given by the polarization vectors ${\vec\varepsilon}_{\lambda}({\vec k})$ which have c-number components. 
Here we consider only  the {\em circular} polarization \cite{BDR} with
$\vec{\varepsilon}_{\pm 1}(\vec{k})=\frac{1}{\sqrt{2}}(\pm
\vec{e}_1+i\vec{e}_2)$, in a three-dimensional orthogonal local
frame $\{\vec{e}_i\}$ where $\vec{k}=k\vec{e}_3$.

\section{Modified Bessel functions}

According to the general properties of the modified Bessel functions, $I_{\nu}(z)$ and $K_{\nu}(z)=K_{-\nu}(z)$ \cite{NIST}, we
deduce that those used here, $K_{\nu_{\pm}}(z)$, with
$\nu_{\pm}=\frac{1}{2}\pm i \mu$ are related among themselves through
\begin{equation}
H^{(1,2)}_{\nu}(z)=\mp\frac{2i}{\pi}e^{\mp \frac{i}{2}\pi\nu}K_{\nu}(\mp iz)\,, \quad z\in {\Bbb R}\,.
\end{equation} 
The functions  used here, $K_{\nu_{\pm}}(z)$ with
$\nu_{\pm}=\frac{1}{2}\pm i \mu$  ($ \mu \in {\Bbb R}$), are related among themselves through
\begin{equation}\label{H1}
[K_{\nu_{\pm}}(z)]^{*}
=K_{\nu_{\mp}}(z^*)\,,\quad \forall z \in{\Bbb C}\,,
\end{equation}
satisfy the equations
\begin{equation}\label{H2}
\left(\frac{d}{dz}+\frac{\nu_{\pm}}{z}\right)K_{\nu_{\pm}}(z)=-K_{\nu_{\mp}}(z)\,,
\end{equation}
and the identities
\begin{equation}\label{H3}
K_{\nu_{\pm}}(z)K_{\nu_{\mp}}(-z)+ K_{\nu_{\pm}}(-z)K_{\nu_{\mp}}(z)=\frac{i\pi}{ z}\,,
\end{equation}
that guarantees the correct orthonormalization properties of the fundamental spinors. For 
$z\to \infty$ these functions behave as  \cite{NIST}
\begin{equation}\label{Km0}
I_{\nu}(z) \to \sqrt{\frac{\pi}{2z}}e^{z}\,, \quad K_{\nu}(z) \to K_{\frac{1}{2}}(z)=\sqrt{\frac{\pi}{2z}}e^{-z}\,,
\end{equation} 
regardless the index $\nu$.

Moreover, here we use the integral (6576-4) of Ref. \cite{GR} with $b=\pm a$,   
\begin{eqnarray}
&&\int_{0}^{\infty}dx\,x^{-\lambda}K_{\mu}(ax) K_{\nu}(\pm ax)=\frac{(\pm)^{\nu}\,2^{-2-\lambda} a^{\lambda-1}}{\Gamma(1-\lambda)}\nonumber\\
&&~~~~\times\Gamma\left(\frac{1-\lambda+\mu+\nu}{2}\right)\Gamma\left(\frac{1-\lambda-\mu+\nu}{2}\right)\nonumber\\
&&~~~~\times\Gamma\left(\frac{1-\lambda+\mu-\nu}{2}\right)\Gamma\left(\frac{1-\lambda-\mu-\nu}{2}\right)\,. \label{KKKK}
\end{eqnarray}


\begin{thebibliography}{20}

\bibitem{P1}
L. Parker, {\em Phys. Rev. Lett.} {\bf  21} (1968) 562.
\bibitem{P2}
L. Parker, {\em Phys. Rev.} {\bf 183} (1969) 1057.
\bibitem{S1}
R. U. Sexl and H. K. Urbantke, {\em Phys. Rev.} {\bf 179} (1969) 1247.
\bibitem{P3}
L. Parker, {\em Phys. Rev. D} {\bf 3} (1971) 246.
\bibitem{c2}
J. Audretsch, {\em Nuovo. Cim.} {\bf  B17} (1973) 284. 
\bibitem{c3}
S. G. Mamaev, V. M. Mostepanenko and  A. A. Starobinsky,{\em  Zh. Eksp. Teor. Fiz.} {\bf 70} (1976) 1577 ({\em Sov. Phys. JETP} {\bf 43} (1976)  823).
\bibitem{c4}
V. M. Frolov, S. G. Mamayev and V. M. Mostepanenko,  {\em Phys. Lett. A} {\bf 55} (1976) 389.
\bibitem{c5}
P.D. D’Eath and J.J. Halliwell, {\em Phys. Rev. D} {\bf 35} (1987) 1100.
\bibitem{Lot1}
K.-H. Lotze, {\em Class. Quant. Grav.} {\bf 4} (1987) 1437.
\bibitem{Lot2}
K.-H. Lotze, {\em Class. Quantum Grav.} {\bf 5} (1988) 595.
\bibitem{Lot3}
K.-H. Lotze, {\em Nuclear Physics B} {\bf 312} (1989) 673.
\bibitem{R1}
I. L. Buchbinder, E. S. Fradkin and D. M. Gitman, {\em Forstchr. Phys.} {\bf 29} (1981) 187.
\bibitem{R2}
I. L. Buchbinder and  L. I. Tsaregorodtsev, {\em Int. J. Mod. Phys A} {\bf 7} (1992) 2055.
\bibitem{R3}
L. I. Tsaregorodtsev, {\em Russian Phys. Journal} {\bf 41} (1989) 1028.
\bibitem{AuSp1}
J. Audretsch and P. Spangehl, {\em Class. Quant. Grav.} {\bf 2} (1985) 733
\bibitem{AuSp2}
J. Audretsch and P. Spangehl, {\em Phys. Rev. D} {\bf 33} (1986) 997.
\bibitem{CQED}
I. I. Cot\u aescu and C. Crucean, {\em Phys. Rev. D} {\bf 87} (2013) 044016.
\bibitem{CIR}
I. I. Cot\u aescu, {\em Eur. Phys. J. C}  {\bf 78} (2019)  769.  
\bibitem{CIR1}
I. I. Cot\u aescu, {\em Int. J. Mod. Phys. A} {\bf 34}  (2019) 1950024.
\bibitem{BD}
N. D. Birrel and P. C. W. Davies,  {\em Quantum Fields in Curved
Space} (Cambridge University Press, Cambridge 1982).
\bibitem{Max}
I. I. Cot\u aescu and C. Crucean, {\em Prog. Theor. Phys.} {\bf 124} (2010) 1051.
\bibitem{Yeni}
D. R. Yennie, D. G. Ravenhall, and R. N. Wilson, {\em Phys. Rev.} {\bf 95}, 500 (1954).
\bibitem{S3}
S. Dolan, C. Doran and A. Lasenby, {\em Phys. Rev. D} {\bf 74} (2006) 064005.
\bibitem{CCS}
 I.I. Cot\u aescu, C. Crucean and C.A. Sporea, {\em Eur. Phys. J. C} {\bf 76} (2016) 102.
 \bibitem{CCS1}
 I.I. Cot\u aescu, C. Crucean and C.A. Sporea, {\em Eur. Phys. J. C}  {\bf 76} (2016) 423.
\bibitem{CdS}
I. I. Cot\u aescu, {\em Phys. Rev. D} {\bf 65}  (2002) 084008.
\bibitem{BDR}
S. Drell and J. D. Bjorken, {\em Relativistic Quantum Fields} (Me
Graw-Hill Book Co., New York 1965).
\bibitem{LL}
V. B. Berestetski, E. M. Lifshitz and L. P. Pitaevski, {\em Quantum Electrodynamics} (Pergamon Press, Oxford 1982).
\bibitem{TH}
B. Thaller,  {\it The Dirac Equation}, (Springer Verlag, Berlin Heidelberg, 1992).
\bibitem{NIST}
F. W. J. Olver, D. W. Lozier, R. F. Boisvert and C. W. Clark, {\em NIST Handbook of Mathematical Functions} (Cambridge University Press, 2010).
\bibitem{GR}
I. S. Gradshteyn and I. M. Ryzhik, {\em Table of Integrals, Series, and Products} (Academic Press, New York 2007).



\end{thebibliography}
\end{document}